# Two Displacive Ferroelectric Phase Transitions in Multiferroic Quadruple Perovskite $BiMn_7O_{12}$

A. Maia[1], M. Kempa[1], V. Bovtun[1], R. Vilarinho[2], C. Kadlec[1], J. Agostinho Moreira[2], A. A. Belik[3], P. Proschek[4], S. Kamba[1]*

[1]*Institute of Physics of the Czech Academy of Sciences, Na Slovance 2, 182 00 Prague 8, Czech Republic*

[2]*IFIMUP, Physics and Astronomy Department, Faculty of Sciences, University of Porto, Rua do Campo Alegre 687, s/n- 4169-007 Porto, Portugal*

[3]*Research Center for Materials Nanoarchitectonics (MANA), National Institute for Materials Science (NIMS), Namiki 1-1, Tsukuba, Ibaraki 305-0044, Japan*

[4]*Faculty of Mathematics and Physics, Charles University, Ke Karlovu 5, 121 16 Prague, Czech Republic*

**e-mail: kamba@fzu.cz*

## Abstract

We report on the microwave, terahertz (THz), infrared and Raman spectroscopic studies of $BiMn_7O_{12}$ ceramics, shedding more light into the nature of two structural phase transitions and their possible relation with ferroelectricity in this compound. We observed a softening of one polar phonon in the THz range on cooling towards 460 and 300 K, i.e., temperatures at which $BiMn_7O_{12}$ undergoes subsequent structural phase transitions from monoclinic $I2/m$ to polar monoclinic $Im$ and triclinic $P1$ phases. The soft phonon causes dielectric anomalies typical for displacive ferroelectric phase transitions. Microwave measurements performed at 5.8 GHz up to 400 K qualitatively confirmed not only the dielectric anomaly at 300 K, but also revealed two other weak dielectric anomalies near the magnetic phase transitions at 60 K and 28 K. This evidences the multiferroic nature of the low-temperature phases, although the relatively high conductivity in the kHz and Hz spectral range prevented us from directly measuring the permittivity and ferroelectric polarization. Some Raman modes sense the magnetic phase transitions occurring near 60 and 25 K, showing that spin-phonon coupling is relevant in this compound and in this temperature range. The deviation of the Mn-O stretching mode frequency from the anharmonic temperature behavior was successfully explained by the spin correlation function calculated from the magnetic contribution to the specific heat.

## 1. Introduction

A-site ordered quadruple perovskites, $AA'_3B_4O_{12}$, exhibit several interesting physical and chemical properties [1,2], such as reentrant structural transitions, [3], inter-site charge transfer and disproportionation [2], giant dielectric constant, [4,5] multiferroicity [6,7] and high catalytic activity. [8]. $AA'_3B_4O_{12}$ has a 12-fold-coordinated A-site and a square-planar-coordinated A′-site, while B-sites have the usual octahedral coordination for perovskites. For manganese, it is possible to have A′ = B, and the composition can be written as $AMn_7O_{12}$ in brief. These manganites can have spin, orbital, and charge degrees of freedom depending on the oxidation state of the A cation [1,3,6,9–12].

The quadruple perovskite $BiMn_7O_{12}$ exhibits three structural and two magnetic phase transitions [13]. Above $T_1$ = 608 K, $BiMn_7O_{12}$ crystallizes in a parent cubic structure, with space group $Im\bar{3}$. Between 460 and 608 K, $BiMn_7O_{12}$ adopts a monoclinic symmetry, with pseudo-orthorhombic metrics (denoted as *I2/m(o)*), and orbital order appears below $T_1$. At $T_2$ = 460 K, $BiMn_7O_{12}$ undergoes a phase transition into a polar monoclinic structure, described by the *Im* space group. Finally, at $T_3$ = 290 K, a triclinic distortion takes place and $BiMn_7O_{12}$ transits into another polar structure, described by the *P*1 space group (assigned as *I*1 in reference [14]). Structural analyses of $BiMn_7O_{12}$ are challenging because of severe domain twinning in single crystals, and anisotropic broadening and diffuse scattering in powder [13]. However, first-principles calculations confirm that non-centrosymmetric structures are more stable than centrosymmetric ones. [13] The energy difference between the *Im* and *P*1 models is very small, and this fact can explain why the *Im* to *P*1 transition is very gradual, and there are no Differential Scanning Calorimetry (DSC) anomalies associated with this transition. [13]. The crystal structure and phase sequence of $BiMn_7O_{12}$ are illustrated in Figure 1.

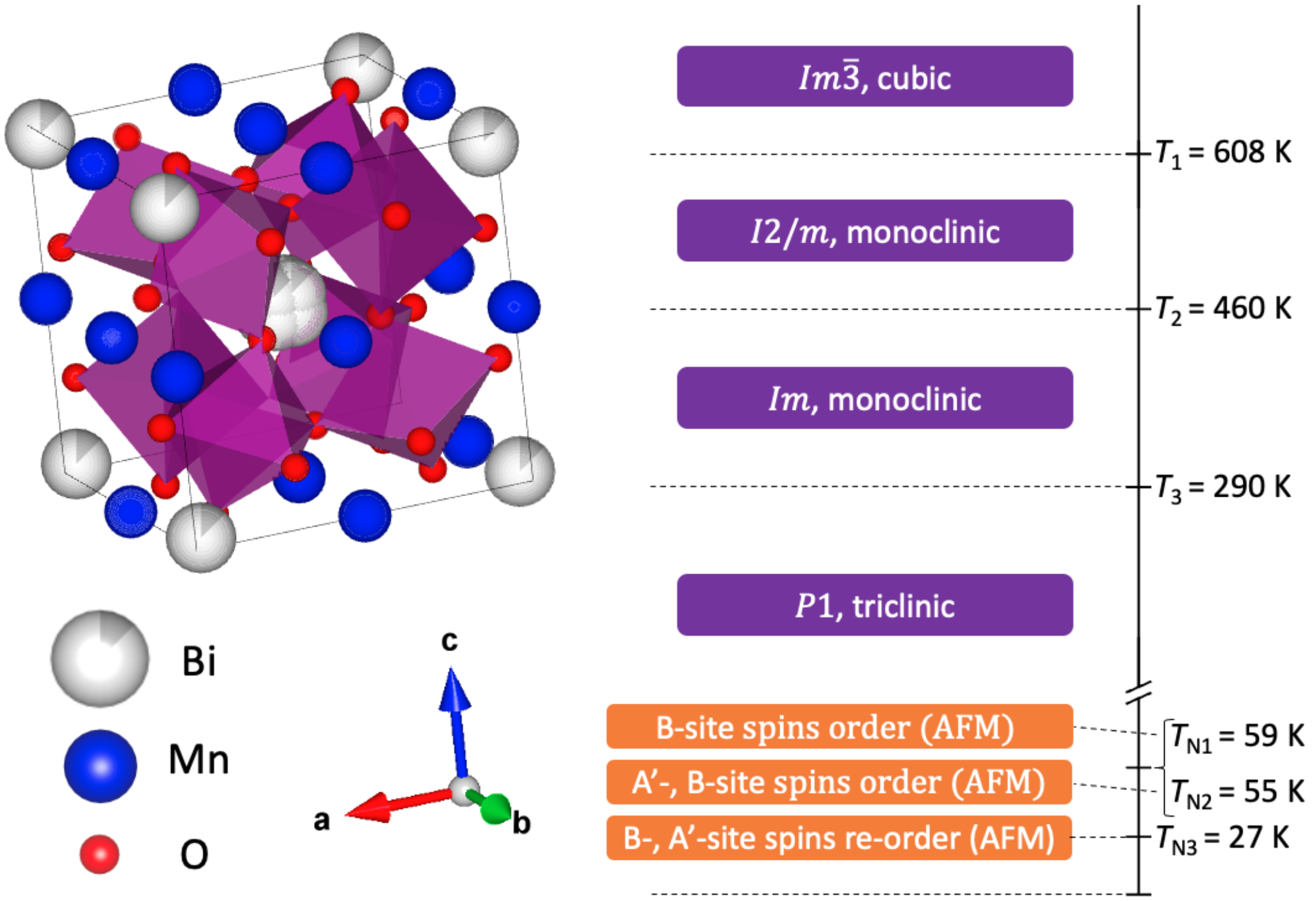

**Figure 1.** Crystal structure of $BiMn_7O_{12}$ at room temperature and schematic representation of its structural (purple) and magnetic (orange) phase sequence. The illustration of the crystal structure was obtained using the VESTA software [15].

In earlier studies, $BiMn_7O_{12}$ has been shown to exhibit two magnetic phase transitions at 55 and 25 K [16,17]. Recent magnetic and powder neutron diffraction experiments revealed three successive magnetic phase transitions [18]. Below $T_{N1}$ = 59 K, B-site Mn cations order antiferromagnetically with propagation vector $\boldsymbol{k}_1$ = (0.5, 0, -0.5), and at $T_{N2}$ = 55 K the A'-site Mn spins order ferrimagnetically with $\boldsymbol{k}_2$ = (0, 0, 0) [18]. Both magnetic orderings coincide down to $T_{N3}$= 27 K, below which both the A'-site and B-site spins reorient with modulation vector $\boldsymbol{k}_3$ = (0, 1, 0). It should be noted that, in contrast to other type II multiferroics, the polar distortion of the lattice in the $P1$ structure of $BiMn_7O_{12}$ stabilizes the E-type magnetic ordering of Mn at the B perovskite positions due to trilinear coupling of two magnetic order parameters $\eta_1$, $\eta_2$ and polarization $P$, i.e., $\eta_1\eta_2P$ [18]. Conversely, in $BiMn_3Cr_4O_{12}$, a different mechanism should be responsible for stabilizing the magnetic structure (probably with a quadratic coupling $\eta^2P$ or $\eta P^2$), because in this material only the Cr spins order below $T_N$ = $T_{FE}$ [7]. Magnetoelectric coupling should be induced in both materials via inverse exchange striction ($\propto \boldsymbol{S}_i \cdot \boldsymbol{S}_j$). The magnetic transitions and the crystal structure are depicted in Figure 1 [16–18].

The relation between the local crystal structure and the spontaneous electric polarization in $BiMn_{6.96}Fe_{0.04}O_{12}$ was recently studied using $^{57}$Fe probe Mössbauer spectroscopy [19,20]. A "dynamic" Born charge model was used to develop an algorithm to construct the temperature dependence of the polarization of the crystal using structural data of the compound and the experimental values of the quadrupole splittings of the Mössbauer spectra of the $^{57}Fe$ probe atoms [19]. The temperature dependence of the electric polarization obtained from the Mössbauer data points out for a paraelectric to ferroelectric first-order phase transition with $T_C \approx 437$ K, close to the *I2/m* to *Im* structural transition temperature $T_2$, as well as, a considerable increase in the electric polarization below 270 K, with extrapolated $T_C \approx 294$ K, close to the *Im* to *P1* structural transition at $T_3$ [19]. It has also been demonstrated that light hole doping of $BiMn_7O_{12}$ with Cu ($BiCu_{0.1}Mn_{6.9}O_{12}$.) can induce incommensurate helical ordering of electric dipoles [21].

Despite $BiMn_7O_{12}$ structural and magnetic properties being extensively studied in the recent past, its dielectric and possible ferroelectric properties in polar phases near $T_2$ and $T_3$ critical temperatures have not yet been published due to its relatively high conductivity above 100 K. $BiMn_7O_{12}$ is sufficiently resistive to perform permittivity measurements only below 90 K, revealing dielectric anomalies at magnetic phase transitions [22,23]. This evidences for magnetodielectric coupling and the possible multiferroic nature of the low temperature magnetic phases. However, electric polarization measurements have not yet been published due to the impossibility of applying a sufficiently high electric field in the measurement of ferroelectric hysteresis loops, and of poling the sample in the paraelectric phase for pyroelectric current studies due to its high conductivity.

Spectroscopic methods that are insensitive to the conductivity of the sample are used to uncover the nature of structural and possible ferroelectric phase transitions in $BiMn_7O_{12}$. That is why we have undertaken a comprehensive temperature dependent infrared, THz, Raman and microwave measurements in a broad temperature range, revealing a ferroelectric soft mode driving both ferroelectric phase transitions near 460 and 290 K, and optical phonons showing anomalies near magnetic phase transitions.

## 2. Experimental details

$BiMn_7O_{12}$ polycrystalline samples were synthesized under high-pressure and high-temperature conditions from stoichiometric mixtures of $Bi_2O_3$ and $Mn_2O_3$ starting reagents, as detailed in reference [13]. Laboratory powder X-ray diffraction (XRD) data were taken at room temperature using a MiniFlex600 diffractometer with $CuK_\alpha$-radiation (2θ range of 10-80°, a step width of 0.02°, and a counting speed of 1 °min-1). XRD data were analyzed by the Rietveld method with RIETAN-2000 program [24]. Weight fractions of impurities were estimated by RIETAN-2000 from refined scale factors. The best ceramics contained some impurities, specifically 1% of $Bi_2O_2CO_3$ and 1% $Mn_2O_3$. The ceramic discs with diameter 6 mm had thickness 1-2 mm. Scanning electron microscope images of $BiMn_7O_{12}$ ceramics revealed grains with the size 2-5 μm (see Figure S1 in Supplementary Material) [25]. The microwave response at 5.8 GHz was measured using the composite dielectric resonator method [26,27]. The $TE_{01\delta}$ resonance frequency, quality factor and insertion loss of the base cylindrical dielectric resonator with and without the sample were recorded during heating from 10 to 400 K with a temperature rate of 0.5 K/min in a Janis closed-cycle He cryostat. The sample without electrodes (2x2 mm plate, 0.32 mm thick) was placed on top of the base dielectric resonator. The resonators were measured in the cylindrical shielding cavity using the transmission setup with a weak coupling by an Agilent E8364B network analyzer. The complex permittivity of the sample(s) was calculated from the acquired resonance frequencies and quality factors of the base and composite resonators.

The magnetic properties were measured via VSM (Vibrating Sample Magnetometer - Quantum Design) using a Quantum Design PPMS (Physical Properties measurement System) in various temperature ranges (down to 2 K) and magnetic field (up to 9 T). The heat capacity measurements were performed by using the heat capacity option in the PPMS in various ranges of temperature (down to 2 K).

The THz complex transmittance was measured using a custom-made time-domain spectrometer powered by a Ti:sapphire femtosecond laser with 35-fs-long pulses centered at 800 nm. The system is based on coherent generation and subsequent coherent detection of ultrashort THz transients. The detection scheme consists of an electro-optic sampling of the electric field of the transients within a 1-mm-thick, (110)-oriented ZnTe crystal as a sensor. [28] This allows measuring the time profile of the THz transients transmitted through the sample. The $BiMn_7O_{12}$ sample was highly absorbing in the THz region due to a strong optical soft mode, therefore we glued it on a sapphire substrate and polished it down to 38 μm. The bare sapphire substrate (thickness 0.543 mm) was measured as a reference.

Low-temperature unpolarized IR reflectivity measurements of one-sided optically polished ceramics (thickness ~1.5 mm) were performed using a Bruker IFS-113v Fourier-transform IR spectrometer equipped with a liquid-He-cooled Si bolometer (1.6 K) serving as a detector. For both the THz complex transmittance and IR reflectivity measurements, the temperature control was done through an Oxford Instruments Optistat optical continuous He-flow cryostats with mylar and polyethylene windows, respectively. A commercial high-temperature cell Specac P/N 5850 was used for IR and THz studies above room temperature. The samples were heated in vacuum up to 580 K. We were concerned about heating the samples to higher temperatures to prevent degradation of the samples. The IR spectra were fitted using one Lorentz oscillator for each phonon mode. The complex dielectric function is given by:

$$\varepsilon(\omega) = \varepsilon_\infty + \sum_j \frac{\Delta\varepsilon_j \omega_{0j}^2}{\omega_{0j}^2 - \omega^2 - i\gamma_j\omega} \tag{1}$$

where $\epsilon_\infty$ is the contribution from electronic transitions to the dielectric function, and the j-th phonon is described by an eigenfrequency $\omega_{0j}$, an oscillator strength $\Delta\epsilon_j$, and damping $\gamma_j$. These parameters were fitted so that the reflectivity at normal incidence, given by $R(\omega) = \left|\frac{\sqrt{\varepsilon(\omega)}-1}{\sqrt{\varepsilon(\omega)}+1}\right|^2$, matches the experimental data. The eigenfrequencies $\omega_{0j}$ correspond to the transverse optical phonon frequencies. The high-frequency permittivity, $\varepsilon_\infty$, resulting from electronic absorption processes, was obtained from the room-temperature frequency-independent reflectivity tail above the phonon frequencies and was assumed temperature independent.

Unpolarized Raman spectra were recorded using a Renishaw inVia Qontor spectrometer with a 785 nm linearly polarized diode-pumped laser and an edge filter. Measurements were done at fixed temperatures from 10 to 600 K using a THMS600 Linkam stage cooled by a nitrogen flow down to 80 K and a custom-made closed-cycle helium cryostat down to 10 K. The laser power (2.3 mW) was chosen adequately to prevent heating the sample. The wavenumber at a given temperature, $\omega(T)$, of each Raman mode is obtained by the best fit of the Raman spectra with a sum of damped oscillators [29]:

$$I(\omega,T) = [1 + n(\omega,T)] \sum_j \frac{A_{0j}\Omega_{0j}^2\Gamma_{0j}\omega}{\left(\Omega_{0j}^2 - \omega^2\right)^2 + \Gamma_{0j}^2\omega^2} \tag{2}$$

where $n(\omega,T)$ is the Bose-Einstein factor and $A_{0j}, \Omega_{0j}, \Gamma_{0j}$ are the strength, wavenumber, and damping coefficient of the j-th oscillator, respectively. In the temperature range where no anomalous behavior is observed, the temperature dependence of the wavenumber of the phonon frequencies can be well described by the normal anharmonic temperature effect due to volume contraction as temperature decreases [30]:

$$\omega(T) = \omega_0 + C\left[1 + \frac{2}{e^x - 1}\right] + D\left[1 + \frac{3}{e^y - 1} + \frac{3}{(e^y - 1)^2}\right] \tag{3}$$

with $x \equiv \hbar\omega_0/2k_BT$, $y \equiv \hbar\omega_0/3k_BT$ and where $\omega_0$, C and D are model constants, $\hbar$ is the reduced Planck constant and $k_B$ is the Boltzmann constant.

Deviations to the normal anharmonic temperature effect were interpreted in the basis of the spin-phonon coupling as [31]:

$$\Delta\omega(T) = \omega(T) - \omega_0 \propto \frac{\partial^2 J}{\partial u^2}\langle S_i \cdot S_j\rangle \approx (R_{AFM} - R_{FM})\langle S_i \cdot S_j\rangle \tag{4}$$

where $J$ is the magnetic exchange integral, $u$ is the normal coordinate of the vibrational mode and RFM and RAFM are the second derivatives of the ferro- and antiferromagnetic exchange integrals with respect to the normal coordinate, where the approximation assumes the same spin correlation function, $\langle S_i \cdot S_j\rangle$, for both.

## 3. Results and Discussion

### 3.1 Soft phonon in the THz range and microwave dielectric properties

Figure *2*a and Figure *2*b show the real and imaginary parts of the complex dielectric spectra, $\varepsilon'(\omega)$ and $\varepsilon''(\omega)$ of $BiMn_7O_{12}$, respectively, measured at several fixed temperatures in the THz spectral range. At 580 K, only one polar phonon is observed at 28 $cm^{-1}$. As the temperature decreases towards $T_2$ = 460 K, the phonon frequency shifts towards lower frequencies, while the static permittivity increases, and on further cooling to 380 K, the phonon hardens. This temperature behavior is typical of a polar soft phonon driving a displacive ferroelectric phase transition. Below 380 K, the phonon softens once again down to 300 K. This phonon anomaly at 300 K is in the vicinity of the structural phase transition from the *Im* to *P*1 structure, in which the ferroelectric polarization is predicted to move out of the *ac* plane [18]. Below 300 K, the phonon frequency smoothly increases down to 10 K. This behavior is best seen in the temperature dependence of the soft mode wavenumber and its dielectric strength, depicted in **Chyba! Nenalezen zdroj odkazů.**Figure 3a and Figure 3b, respectively. The dielectric strength of the soft phonon, $\Delta\varepsilon_{SM}(T)$, considerably increases on cooling from 580 K towards 460 K (see Figure 3b) due to the conservation law of the oscillator strength $f_j = \Delta\varepsilon_j(T)\omega_{0j}^2 = const$ (valid for all uncoupled polar phonons, including the soft mode). It is noteworthy that at 580 K, $\Delta\varepsilon_{SM}$ constitutes approximately 90% of the static electric permittivity in the THz range: $\varepsilon_0 = \varepsilon_\infty + \sum_j \Delta\varepsilon_j$. Cooling from 380 K, $\Delta\varepsilon_{SM}(T)$ decreases and exhibits a broad plateau-like anomaly down to 300 K, below which it smoothly decreases down to 20% of the static electric permittivity at the lowest measured temperature.

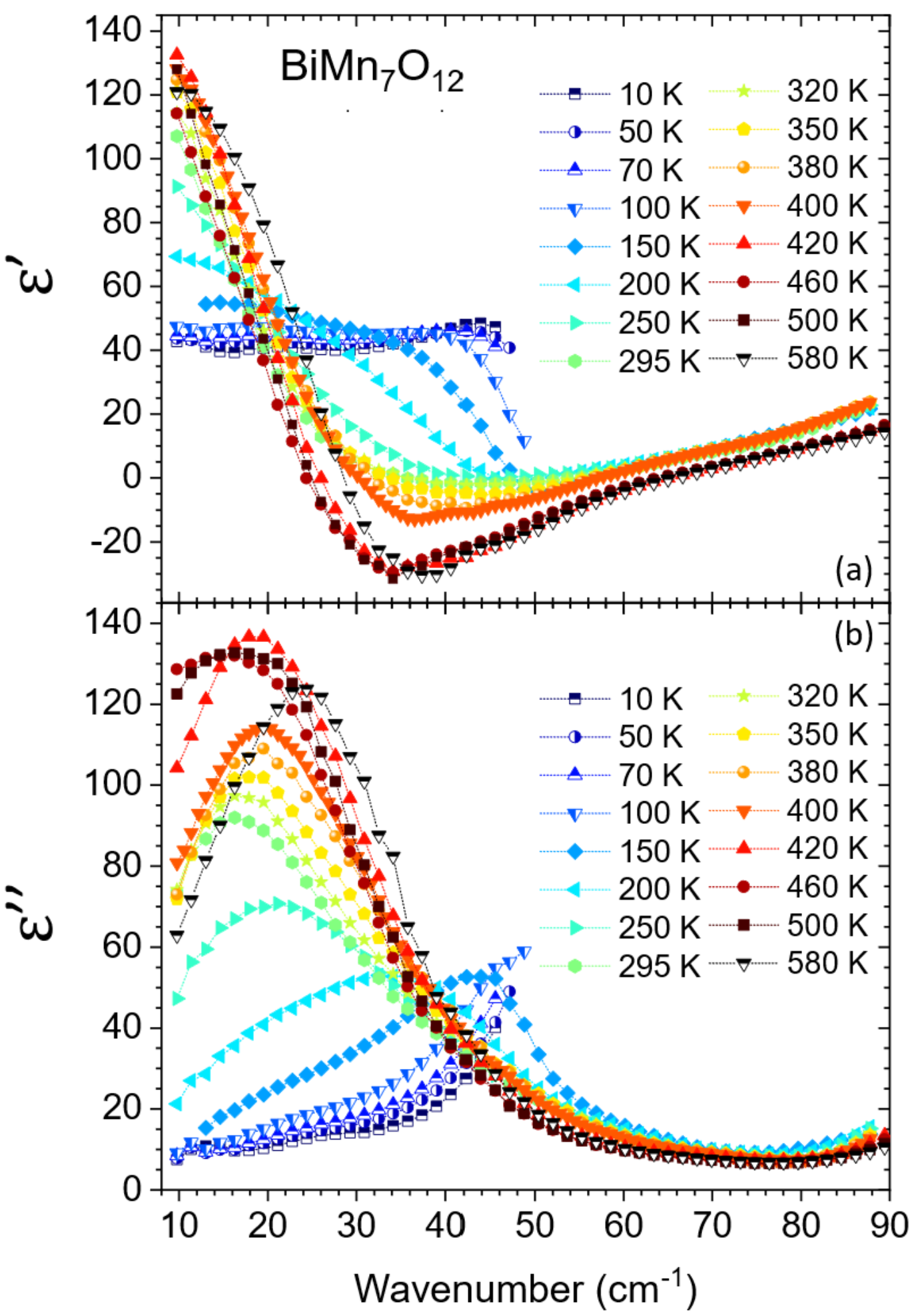

**Figure 2.** (a) $\varepsilon'(\omega)$ and (b) $\varepsilon''(\omega)$ spectra of $BiMn_7O_{12}$ obtained by THz time-domain spectroscopy at several temperatures.

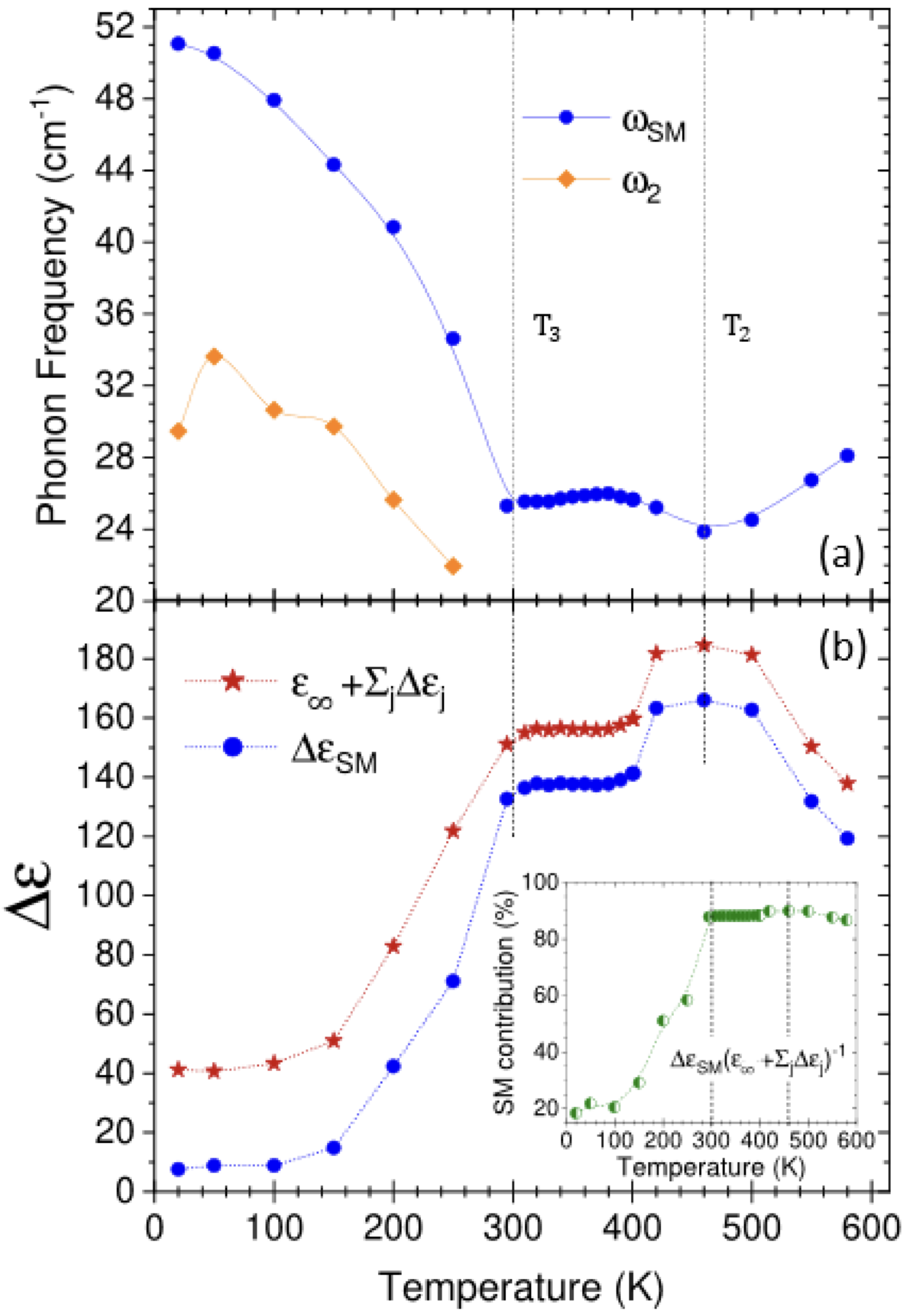

**Figure 3.** (a) Temperature dependence of the soft-mode frequency, $\omega_{SM}$, and newly activated mode below room temperature in $BiMn_7O_{12}$. (b) Temperature dependence of dielectric strength of the soft mode and the sum of the contributions of all phonons to the static permittivity. The dashed lines are a guide for the eye. Inset: Contribution (in %) of the SM to the total static permittivity.

The temperature dependence of the relative changes of real and imaginary parts of the dielectric permittivity measured at 5.8 GHz, is shown in Figure 4a and Figure 4b, respectively. A broad anomaly at 300 K is clearly visible, which is qualitatively consistent with the phonon anomaly in Figure 3 near $T_3$. Its absolute value is smaller than $\varepsilon$ in Figure 3, most likely due to lower accuracy of the MW measurements, because the rectangular sample plate is much smaller than base dielectric resonator while the exact calculation of $\varepsilon$ requires the same diameter of the cylindrical sample and base resonator [26]. The anomalies seen at magnetic phase transitions near 60 and 28

K (see also our magnetic moment measurements in Figure S2 [25]) can come from both magnetic permeability μ and electric permittivity $\varepsilon$, but it should be mentioned that dielectric measurements at 100 kHz revealed similar anomalies in $\varepsilon(T)$ [23] and also in the Hz frequency range (see Figure S3 [25]); so we assume that the dielectric permittivity contributes largely to the anomalous temperature dependence observed in Figure 4 at low temperatures and possible contribution of permeability to anomalies at magnetic phase transitions is negligible. Unfortunately, due to the experimental limitations of our MW setup, it was not possible to measure above 400 K; so the effect of the higher temperature structural transitions ($T_1$ and $T_2$) in the MW permittivity are not ascertained [15]. The origin of the small anomaly around 95 K is unknown, but it should be noted that, at the same temperature, an anomaly in the temperature dependence of the magnetization was observed (see Figure S2 (b) [25]), but no anomaly in specific heat was seen near 95 K (see Figure ).

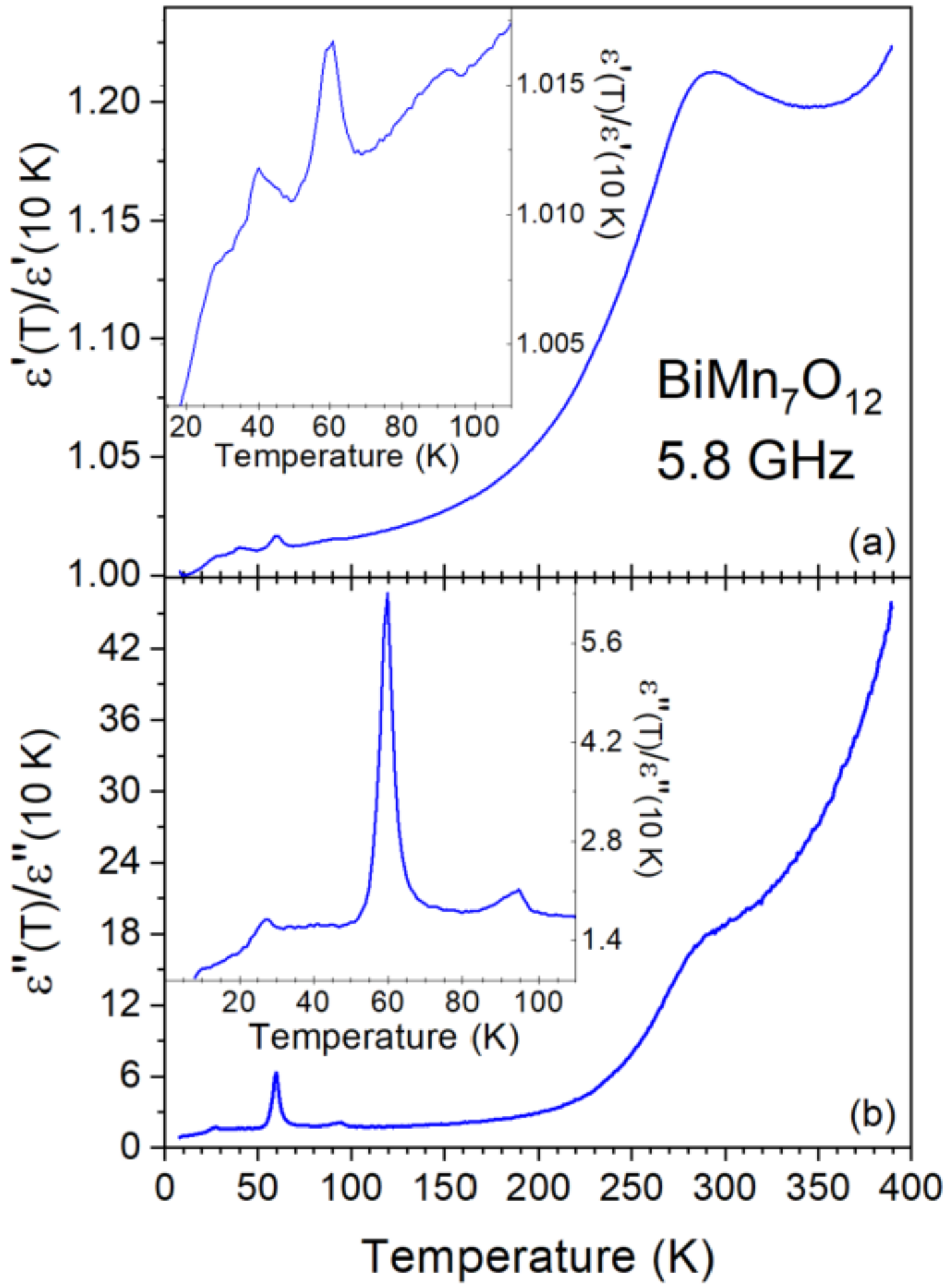

**Figure 4.** Relative changes of the (a) real and (b) imaginary parts of the dielectric permittivity measured at 5.8 GHz.

### 3.2 Polar phonons

To better describe the temperature dependence of the polar soft phonon and its contribution to the total permittivity, we have also probed the higher energy polar lattice excitations through temperature dependent FTIR spectroscopy. Representative unpolarized reflectivity spectra, recorded at 20, 300 and 500 K, are shown in Figure 5. The MIR spectrum at 300 K is shown in Figure S4, and the fit parameters are listed in Table SI [25]. The spectra of the real and imaginary parts of permittivity, calculated from fitting the IR reflectivity spectra using Equation 1 are also shown in Figure 6a and Figure 6b, respectively.

The factor-group analysis and the optical activity of phonons for the different structural phases is presented in Table 1. From symmetry point of view, 18 new polar modes and 36 new Raman modes become active due to a change of symmetry from monoclinic $I2/m$ to the lower symmetry monoclinic $Im$ at 460 K. When changing from $Im$ symmetry to $P1$ at 290 K, no new IR or Raman modes are expected. However, it should be emphasized that in the ferroelectric *Im* and $P$1 phases all phonons are both IR and Raman active.

The significant increase in the number of polar modes with cooling can be seen both in **Figure 5** and **Figure 6**. Experimentally, we detected 41 IR-active modes at 20 K ($P$1 phase), 31 modes at 300 K (*Im* phase), and 28 modes at 500 K (*I2/m* phase). This is below the number of modes expected from the factor-group analysis. Both the high number of modes and the temperature ranges of the phase transitions leading to a higher damping of the modes, explain the lower number of detected modes. Furthermore, many of those modes can have an intensity that is below our detection limit.

**Table 1:** Factor-group analysis of the $\Gamma$-point optical phonons in $BiMn_7O_{12}$ [32–34]. The Wyckoff positions of atoms were taken from reference [13].

| Temperature | Space Group | IR-active Modes | Raman-active Modes | Silent Modes |
|---|---|---|---|---|
| $T > T_1$ | $Im\bar{3}$ | $14T_u$ | $3A_g \oplus 3E_{1g} \oplus 3E_{2g} \oplus 4T_g$ | $3A_u \oplus 3E_{1u} \oplus 3E_{2u}$ |
| $T_2 < T < T_1$ | $I2/m$ | $19A_u \oplus 23B_u$ | $13A_g \oplus 11B_g$ | — |
| $T_3 < T < T_2$ | $Im$ | $33A' \oplus 27A''$ | $33A' \oplus 27A''$ | — |
| $T < T_3$ | $P1$ | $60A$ | $60A$ | — |

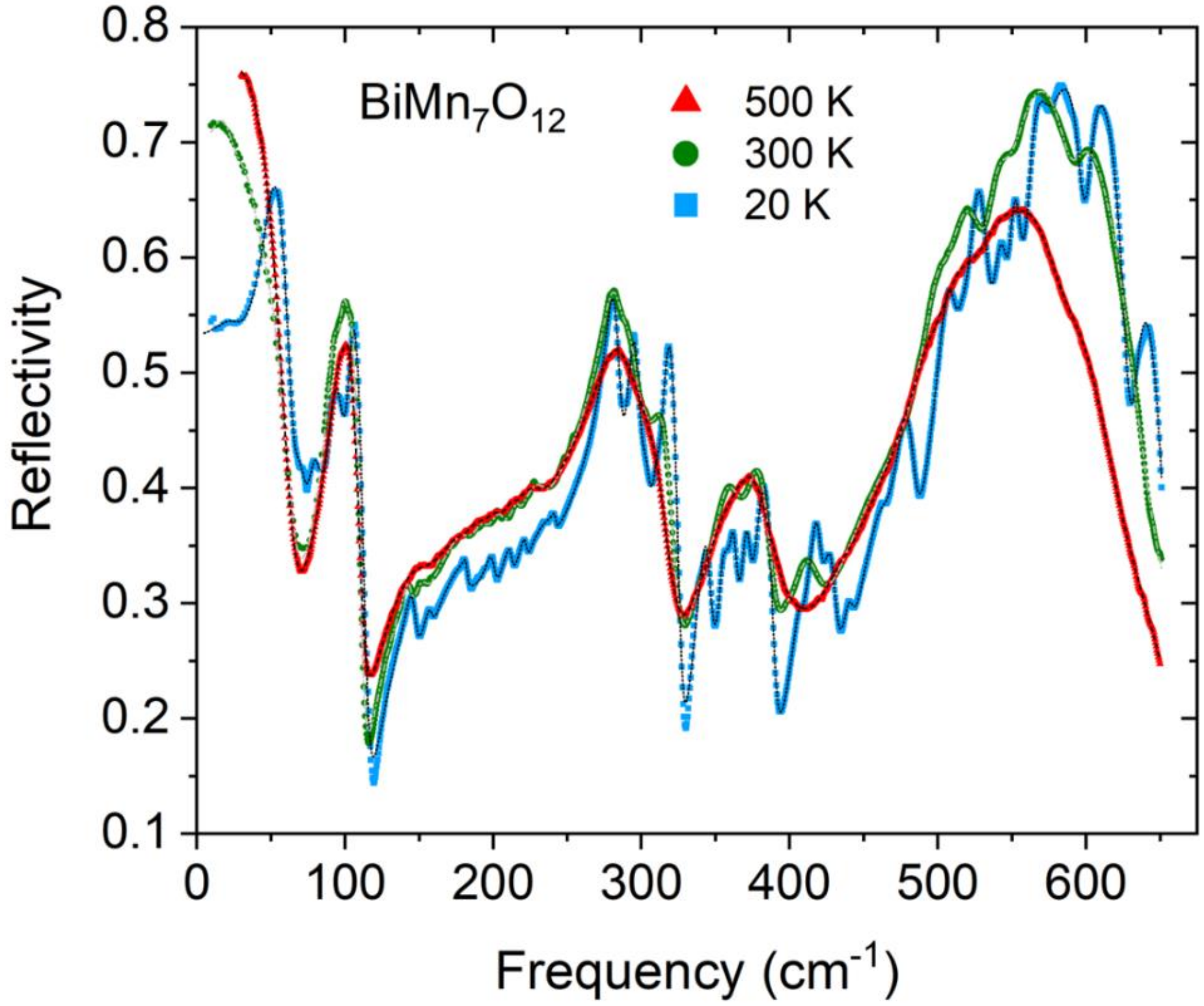

**Figure 5.** Unpolarized IR reflectivity spectra of $BiMn_7O_{12}$ ceramics recorded at 20 K, 300 K, and 500 K. The corresponding fits are shown as dotted lines. The reflection band below 50 $cm^{-1}$ corresponds to the soft mode.

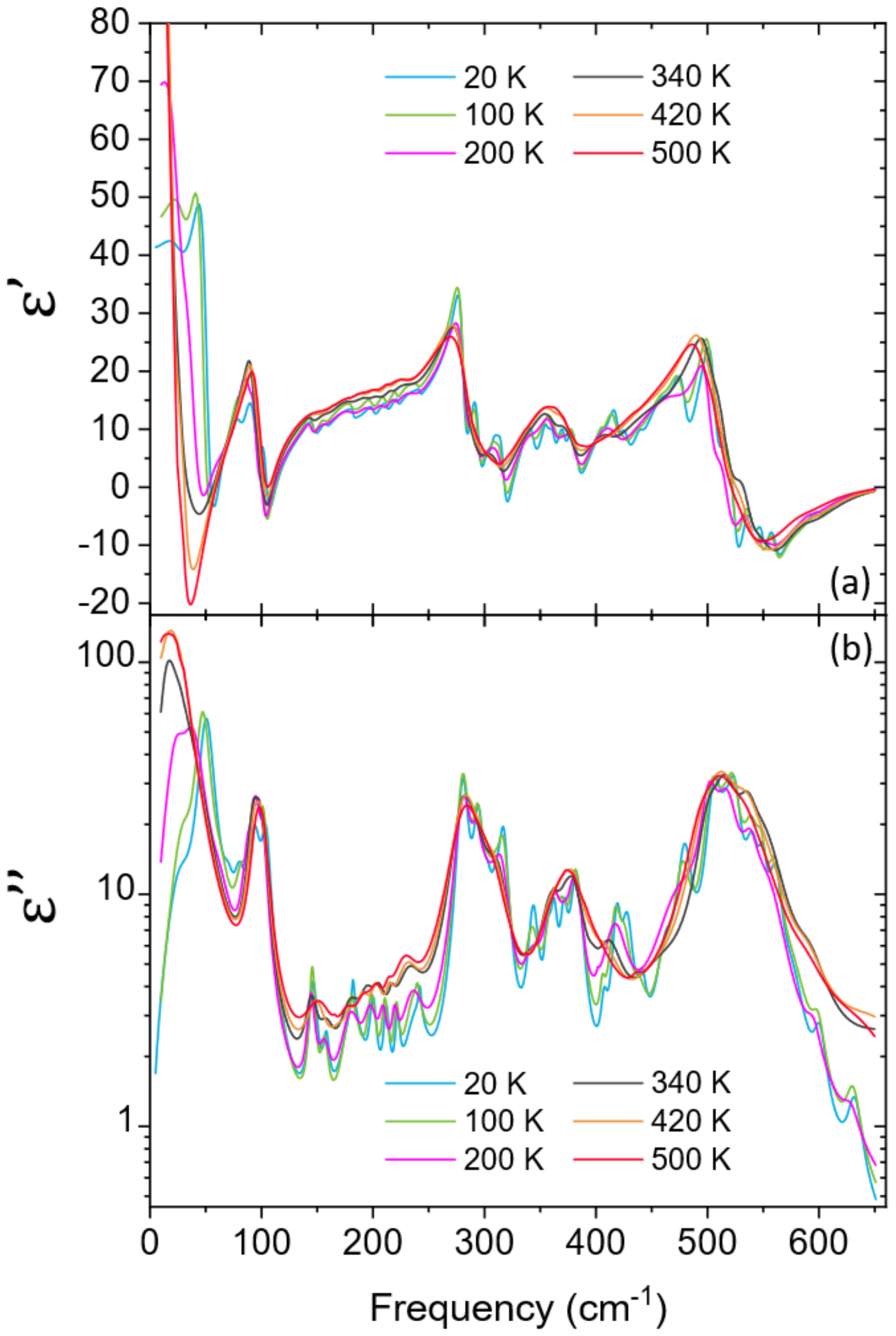

**Figure 6.** (a) $\varepsilon'$ and (b) $\varepsilon''$ (in logarithmic scale) spectra of $BiMn_7O_{12}$ obtained from fitting the IR reflectivity at several temperatures.

### 3.3 Specific heat

The specific heat, divided by temperature, as a function of temperature, is shown in Figure . Two anomalies are clearly observed at 59 K and at 24 K, respectively, which are in good agreement with the reported values of the magnetic phase transitions occurring at $T_{N1}$ = 59 K and at $T_{N3}$ = 23 K, respectively, associated with the paramagnetic to the E-type antiferromagnetic ordering of the B-site $Mn^{3+}$ spins, and a change of spin ordering of both A'- and B-site Mn sublattices (ferrimagnetic moment of Mn in A' sites reduces and antiferromagnetic vector of B-site Mn changes direction). As in previous reports, the specific heat does not reveal any anomalous temperature dependence at $T_{N2}$ nor at the structural phase transitions occurring below 290 K. The analysis of the specific heat versus temperature, by means of the fitting of Debye´s equation to the experimental data, enabled us to calculate the magnetic contribution to the specific heat, shown in Figure  (neglecting the electronic contribution only relevant below 10 K) [25]. The magnetic contribution to the specific heat increases just below 150 K, due to precursor magnetic effects. The existence of precursor effects well above the magnetic phase transition has also been observed in rare-earth orthomanganites, like $GdMnO_3$ [35].

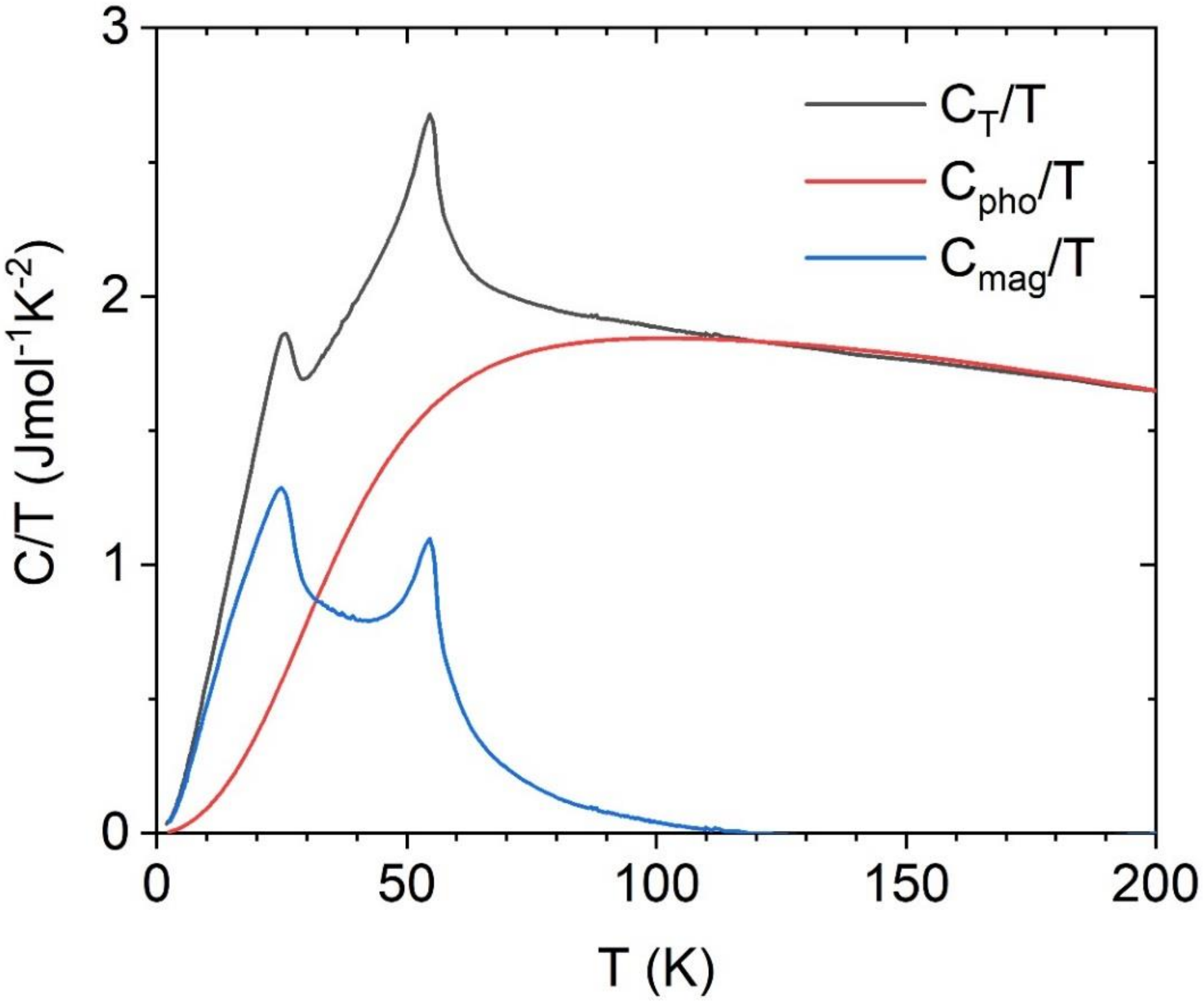

**Figure 7.** Temperature dependence of the specific heat divided by temperature of $BiMn_7O_{12}$. Solid red line was determined by the best fit of Debye equation to the experimental data recorded above 120 K. Temperature dependence of the magnetic contribution to the specific heat, calculated from the difference of the experimental data and the extrapolated lattice contribution described by Debye behavior, extrapolated to 2 K.

### 3.4 Raman scattering and spin-phonon coupling

Figure  shows the unpolarized Raman spectra recorded at different fixed temperatures, in the 100 – 850 $cm^{-1}$ spectral range. The high temperature Raman spectra reveal rather broad bands, due to thermal effects. The Raman spectrum recorded at 600 K is properly simulated with 7 bands in this spectral range. According to group theory arguments and published crystallographic data [13], in the $Im\overline{3}$ cubic phase only phonons involving vibrations of oxygen atoms are Raman active. As expected, on cooling, the Raman bands become narrower and shift towards higher wavenumbers. The phase transition at $T_1$ = 608 K impacts certain spectral features, with the most evident

being the splitting of the band located at 175 $cm^{-1}$ and at 633 $cm^{-1}$, at 600 K, and the appearance of sharp bands at 158 $cm^{-1}$ in the spectrum recorded at 380 K. The spectrum recorded at 400 K is described by 12 bands. The structural phase transition from *Im* to *P1*, at 290 K is revealed by the appearance of weak bands, better observed in the 470 – 550 $cm^{-1}$. However, the number of new bands is smaller than the number of predicted new modes in the Raman spectra recorded in the monoclinic and triclinic phases; this discrepancy is explained by a weak intensity of modes and partial/total band overlap. Nevertheless, a detailed analysis of the temperature dependencies of the frequency of some phonons reveals interesting results, which we will address in the following.

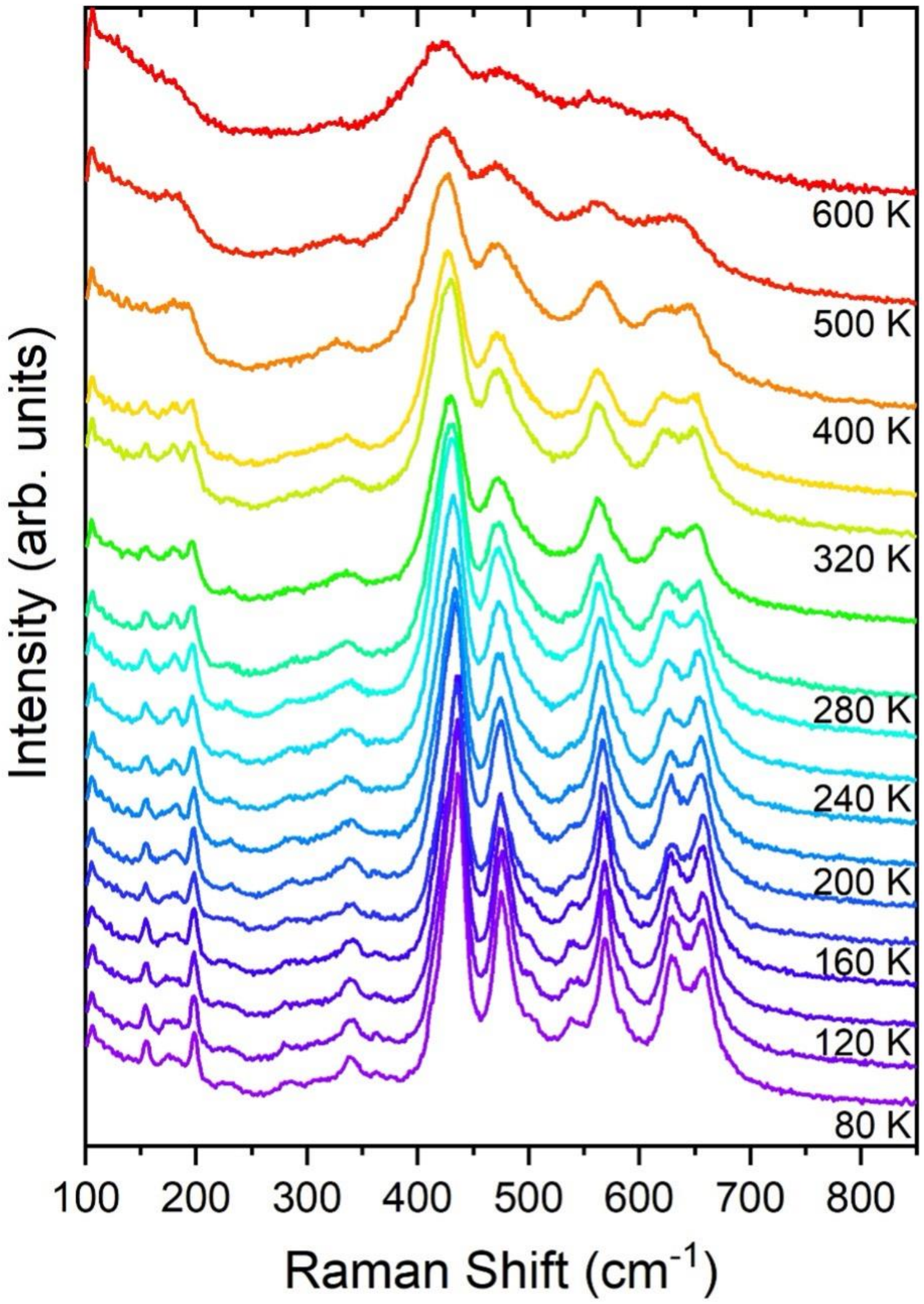

**Figure 8.** Unpolarized Raman spectra of $BiMn_7O_{12}$ ceramics recorded at several fixed temperatures, in the 100 to 850 $cm^{-1}$ spectral range. The spectra are vertically offset from each other for better resolution.

Figure 7(a) shows the temperature dependence of the phonon frequency, observed at 153 $cm^{-1}$ at 440 K. This phonon is only observed below $T_2$ = 460 K, and its frequency exhibits a cusp like anomaly at ~ 294 K, exactly at the *Im* to *P*1 structural phase transition. In rare-earth orthomanganites and rare-earth orthoferrites, low frequency phonons lying in this spectral range, are assigned to the A-site atomic vibrations. The appearance of this band at the cubic ($Im\bar{3}$) to the monoclinic ($I2/m$) phase transition at $T_2$ reveals the Raman activation of vibrations involving other than just oxygen movements. Although no mode assignment is available to the best of our knowledge, we tentatively assign this phonon to the A-site vibration, likely associated with $Bi^{3+}$ oscillations, which is expected to participate on the soft mode, already evidenced in THz spectroscopic studies, and the stabilization of the ferroelectric phase. This assignment is also supported by the current interpretation of the Raman spectra recorded in $LaMn_7O_{12}$ [36].

The magnetic phase transition occurring at $T_{N1}$ = 59 K is also reflected by the anomalous temperature dependence of some internal vibrations. As a representative example, we present in Figure 7(a) and Figure 7(b) the temperature dependence of the modes at 559 $cm^{-1}$ (value at 400 K) and 633 $cm^{-1}$ (value at 600 K), respectively. These two modes are known to involve the oxygen vibrations and, consequently, they account for the changes in bond angle O-Mn-O and length Mn-O, respectively. Therefore, these vibrations probe the magnetic exchange interactions. On cooling, the wavenumber of these two modes increases, following the anharmonic temperature behavior described by the solid lines, determined by the fit of Equation 3 to the experimental data above 200 K and extrapolated to low temperatures. No clear anomalies are ascertained in the structural phase transitions, but a detailed inspection of the temperature dependence of the wavenumber of these two modes shows a clear deviation from the extrapolated anharmonic thermal behavior. The deviation, regarding the extrapolated anharmonic temperature behavior of the wavenumber of the two modes, is clear at $T_{N1}$ = 59 K, for further cooling, an anomalous increase is observed, due to spin-phonon coupling in this compound. As representative example, we focus on the Mn-O vibration at ~570 $cm^{-1}$ (value at 5 K), shown in Figure 7(a)), which better senses the magnetic phase transition at $T_{N1}$. According to spin-phonon coupling theory presented in Equation 4, the anomalous temperature dependence of the wavenumber must be a linear function of the spin-spin correlation function, $\langle S_i \cdot S_j \rangle$, which can be calculated from the integral of the magnetic contribution to the specific heat show in Figure 7 [31]. In this case, the anomalous temperature dependence of the wavenumber of the aforementioned Mn-O vibration is a linear function of the spin-spin correlation function, as shown in Figure 7(c), providing clear evidence for the coupling between lattice vibrations and the magnetic momenta in $BiMn_7O_{12}$.

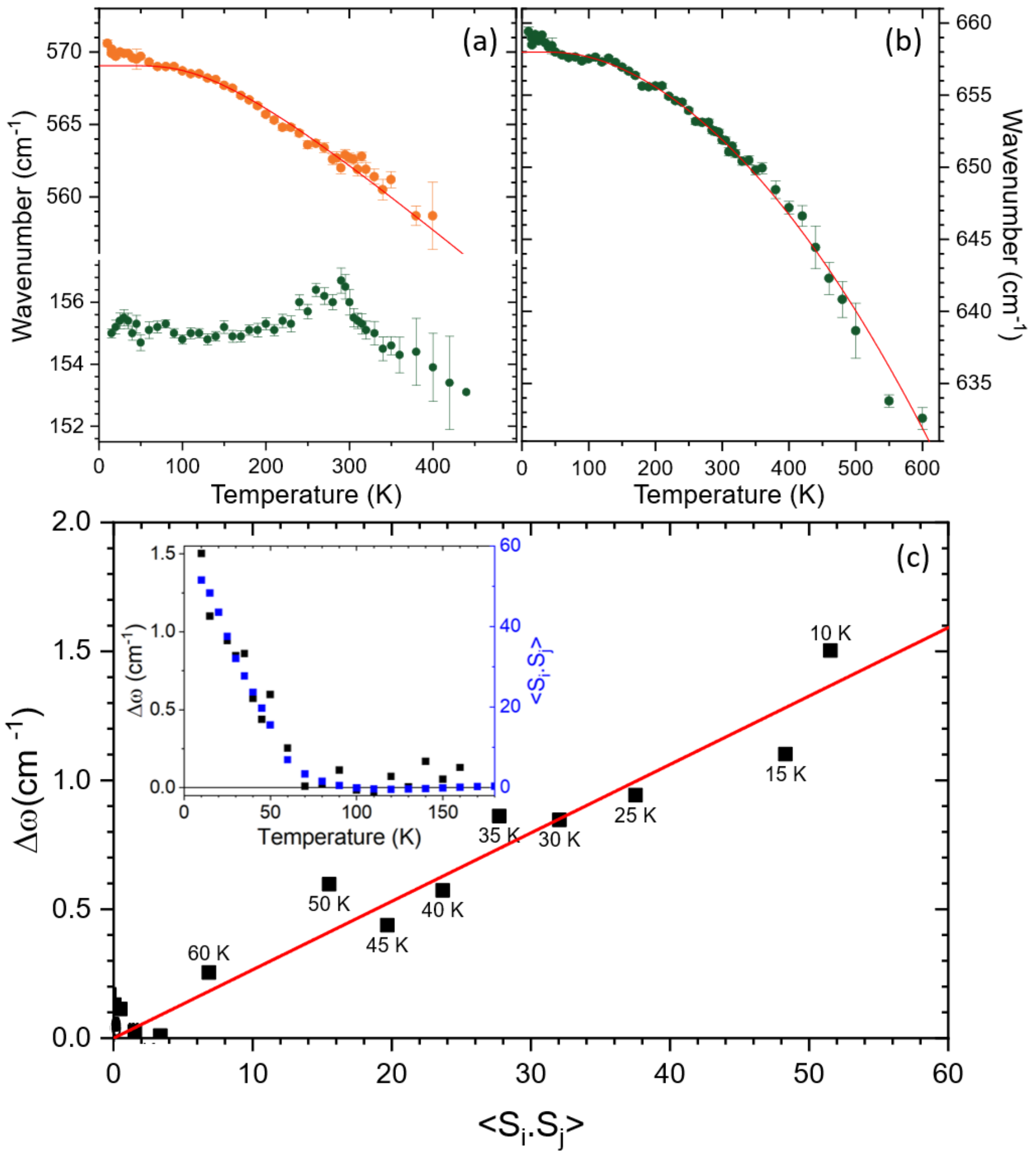

**Figure 7.** (a)-(b) Temperature dependence of the wavenumber of selected Raman modes. The best fit with Equation 3 in the paraelectric phase is represented by a solid line and is extrapolated down to 0 K. (c) The deviation, $\Delta\omega$, of the highest phonon frequency seen near 570 $cm^{-1}$ from the anharmonic temperature behavior (Equation 3) as a function of the spin-spin correlation function, $< S_i \cdot S_j >$, obtained from integrating the magnetic contribution to the specific heat, shown in Figure . Inset: temperature dependencies of $\Delta\omega$ and $< S_i \cdot S_j >$.

## 4. Conclusions

We observed the softening of a phonon in the THz range on cooling towards $T_2 = 460$ K, temperature at which $BiMn_7O_{12}$ undergoes a structural phase transition from monoclinic $I2/m$ to a non-centrosymmetric monoclinic *Im* phase. At 300 K, this phonon also has an anomaly coinciding with another structural phase transition to the $P1$ phase, in which the polarization is predicted to move out of the *ac* plane. The observed behavior of the optical phonon is typical for displacive ferroelectric phase transitions. Also, the microwave permittivity shows a peak at $T_3 = 290$ K, indicating the ferroelectric nature of the phase transition at this temperature. For experimental reasons, it was not possible to measure the microwave permittivity above 400 K, i.e., not near $T_2$ or $T_1$. Unfortunately, neither the polarization hysteresis loops nor the pyrocurrent could be measured due to the conductivity of the sample above 100 K, but the soft mode in THz and IR spectra clearly indicates signs of successive displacive ferroelectric phase transitions in the two polar phases *Im* and $P1$.

The results here presented are consistent with the very recently reported temperature dependence of the polarization calculated from Mössbauer data for $BiMn_{6.96}Fe_{0.04}O_{12}$, showing a paraelectric to ferroelectric phase transition around 437 K, and a considerable increase in polarization below 270 K, with an extrapolated Curie temperature of 294 K [19].

The dielectric anomalies (measured at 5.8 GHz) observed near $T_{N1}$ and $T_{N3}$ are indicative of a strong magnetoelectric coupling at the magnetic phase transitions, for which a phenomenological model of magnetoelectric coupling via inverse exchange-striction has been recently proposed [21].

The Raman mode seen near 155 $cm^{-1}$ senses the structural phase transition at 300 K. Furthermore, some Raman modes sense the magnetic phase transitions occurring at 59, 55 and 27 K, showing that spin-phonon coupling is relevant in this compound in this temperature range. The high-frequency mode observed near 570 $cm^{-1}$ exhibits a clear deviation from anharmonic temperature dependence, which is linearly correlated with the spin correlation function $< S_i \cdot S_j >$.

## Acknowledgements

The authors thank Jan Maňák and Maxim Savinov, respectively, for SEM characterization of the ceramic grains and permittivity measurements in the Hz range below 100 K. This work was supported by the VEGA 2/0137/19 project and the Czech Science Foundation (Project No. 21-06802S).

## Competing interests

The authors declare no competing interests.

# Supplemental Information

# Two displacive Ferroelectric Phase Transitions in Multiferroic Quadruple Perovskite $BiMn_7O_{12}$

A. Maia[1], C. Kadlec[1], M. Kempa[1], V. Bovtun[1], R. Vilarinho[2], J. Agostinho Moreira[2], Alexei A. Belik[3], P. Proschek[4], S. Kamba[1]*

[1]*Institute of Physics of the Czech Academy of Sciences, Na Slovance 2, 182 21 Prague 8, Czech Republic*

[2]*IFIMUP, Physics and Astronomy Department, Faculty of Sciences, University of Porto, Rua do Campo Alegre 687, s/n- 4169-007 Porto, Portugal*

[3]*International Center for Materials Nanoarchitectonics (WPI-MANA), National Instsitute for Materials Science (NIMS), Namiki 1-1, Tsukuba, Ibaraki 305-0044, Japan*

[4]*Faculty of Mathematics and Physics, Charles University, Ke Karlovu 5, 121 16 Prague, Czech Republic*

**e-mail: kamba@fzu.cz*

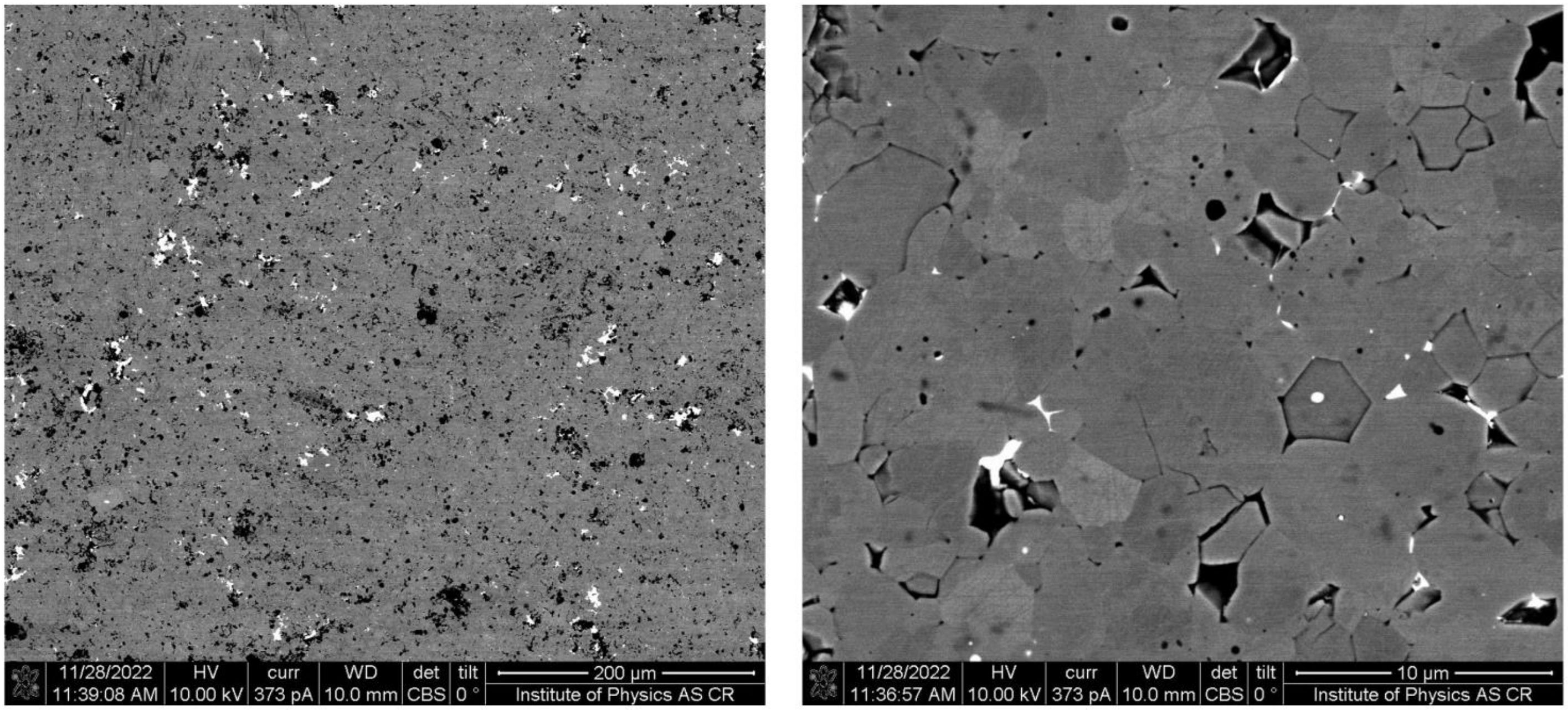

**Figure S1.** Scanning electron microscope (SEM) images of $BiMn_7O_{12}$ ceramics at different scales, indicated in the lower right corner of each panel. Grains with a diameter of several μm are visible. Measured using ThermoFisher FEI Quanta 3D SEM microscope.

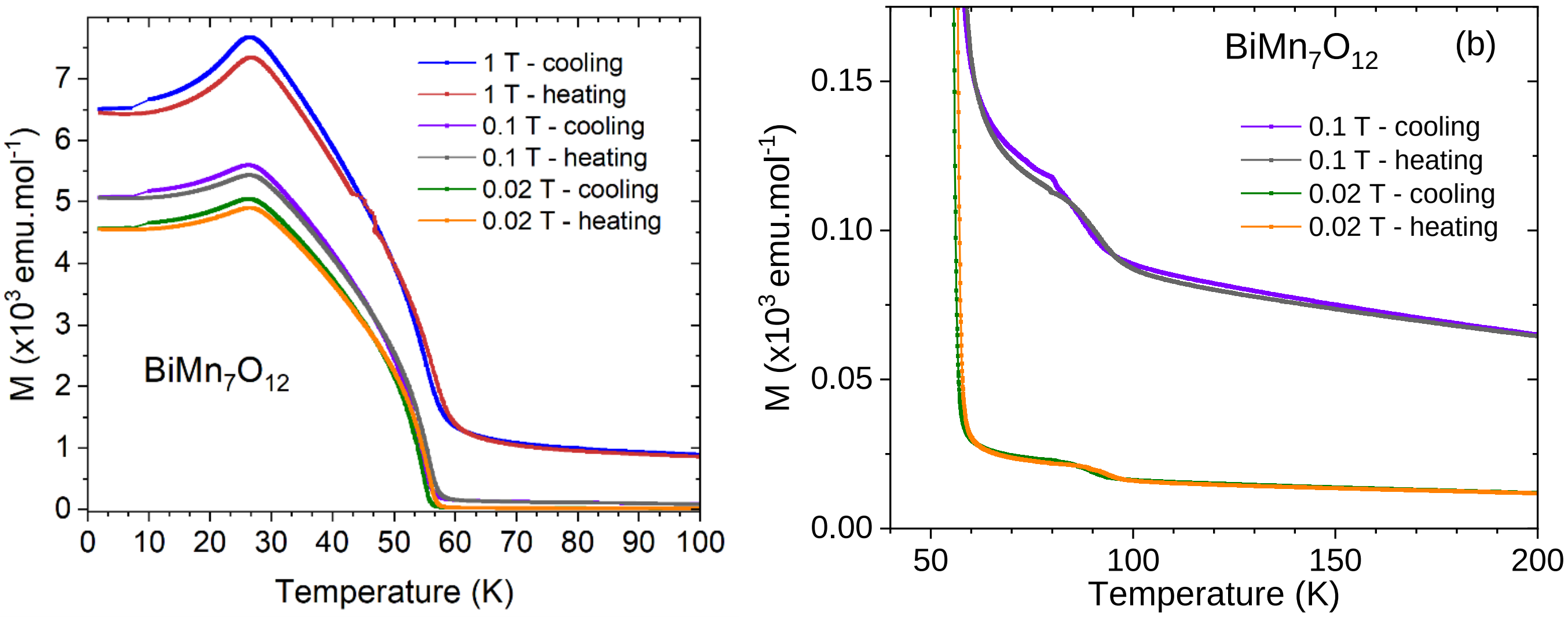

**Figure S2.** (a) Temperature dependence of magnetization of $BiMn_7O_{12}$, measured in heating and cooling runs under several static magnetic fields. Similar magnetic anomalies were published in References [1-3]. (b) Detail of the temperature dependence of magnetization in a weak magnetic field showing a weak anomaly around 90 K. The origin of this anomaly is unknown, but at this temperature the weak dielectric anomaly in Figure 4 (main text) is also visible.

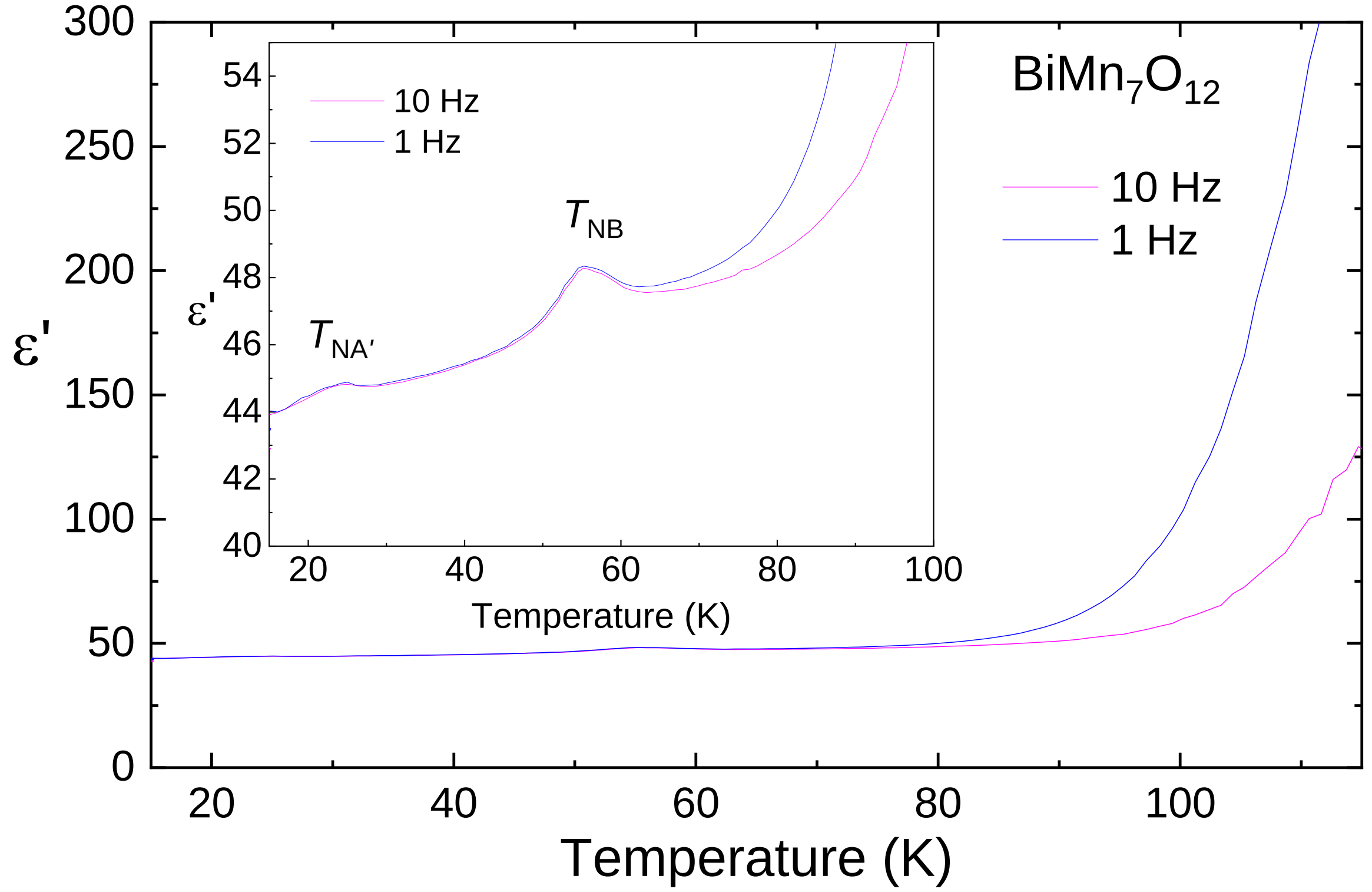

**Figure S3.** Temperature dependence of dielectric permittivity near magnetic phase transitions. The same anomalies are seen in microwave permittivity at 5.8 GHz (see Figure 4 in the main text). The permittivity tends to very large values above 100 K due to inhomogeneous conductivity in the ceramics and grains (Maxwell-Wagner polarization).

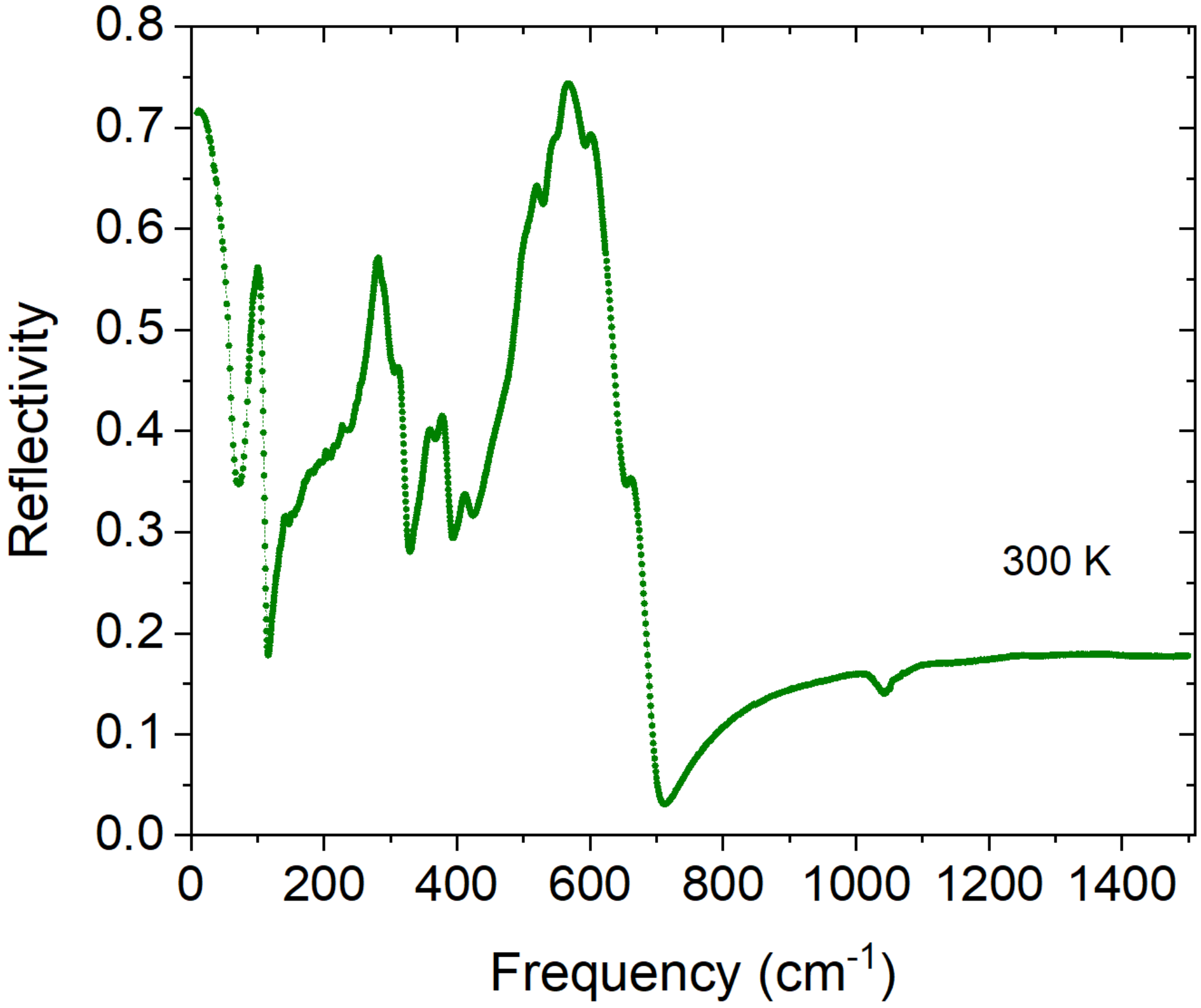

**Figure S4.** Representative example of the IR reflectivity spectra of $BiMn_7O_{12}$ recorded at 300 K, measured up to 1500 $cm^{-1}$. The anomaly around 1050 $cm^{-1}$ is artificially induced by water in the KBr beam splitter used for middle IR measurement.

**Table SI.** Parameters used to fit the IR spectra of $BiMn_7O_{12}$ with the 3-parameter model, at 20 K, 300 K and 500 K. A temperature independent $\epsilon_\infty = 6.80$ was used.

| **Fit Parameters** | **Temperature** | | |
|---|---|---|---|
| | **20 K** | **300 K** | **500 K** |
| $\Delta\epsilon_1$ | 7.57 | 78.52 | -- |
| $\omega_{01}$ (cm$^{-1}$) | 29.5 | 19.5 | -- |
| $\gamma_1$ (cm$^{-1}$) | 27.2 | 20.1 | -- |
| $\Delta\epsilon_2$ | 15.18 | 19.16 | 119.80 |
| $\omega_{02}$ (cm$^{-1}$) | 51.1 | 32.5 | 25.8 |
| $\gamma_2$(cm$^{-1}$) | 14.3 | 22.7 | 25.4 |
| $\Delta\epsilon_3$ | 0.66 | 7.25 | -- |
| $\omega_{03}$ (cm$^{-1}$) | 70.4 | 49.1 | -- |
| $\gamma_3$ (cm$^{-1}$) | 11.6 | 31.2 | -- |
| $\Delta\epsilon_4$ | 0.93 | 1.78 | 2.28 |
| $\omega_{04}$ (cm$^{-1}$) | 80.5 | 93.2 | 96.6 |
| $\gamma_4$ (cm$^{-1}$) | 11.8 | 9.2 | 13.9 |
| $\Delta\epsilon_5$ | 1.99 | 1.00 | 0.84 |
| $\omega_{05}$ (cm$^{-1}$) | 94.0 | 98.6 | 101.5 |
| $\gamma_5$ (cm$^{-1}$) | 12.7 | 7.3 | 10.1 |
| $\Delta\epsilon_6$ | 1.18 | 0.63 | 0.055 |
| $\omega_{06}$ (cm$^{-1}$) | 104.1 | 103.1 | 107.3 |
| $\gamma_6$ (cm$^{-1}$) | 8.7 | 8.6 | 7.2 |
| $\Delta\epsilon_7$ | 0.16 | 0.065 | 0.38 |
| $\omega_{07}$ (cm$^{-1}$) | 146.5 | 144.0 | 152.0 |
| $\gamma_7$ (cm$^{-1}$) | 7.9 | 5.9 | 32.6 |
| $\Delta\epsilon_8$ | 0.045 | 0.078 | 0.0024 |
| $\omega_{08}$ (cm$^{-1}$) | 158.2 | 153.8 | 165.5 |
| $\gamma_8$ (cm$^{-1}$) | 6.3 | 10.5 | 3.8 |
| $\Delta\epsilon_9$ | 0.024 | -- | -- |
| $\omega_{09}$ (cm$^{-1}$) | 182.1 | -- | -- |
| $\gamma_9$ (cm$^{-1}$) | 3.4 | -- | -- |
| $\Delta\epsilon_{10}$ | 0.30 | 0.13 | 0.098 |
| $\omega_{010}$ (cm$^{-1}$) | 182.4 | 181.0 | 177.1 |
| $\gamma_{10}$ (cm$^{-1}$) | 23.7 | 16.8 | 18.7 |
| $\Delta\epsilon_{11}$ | 0.0755 | 0.073 | 0.12 |
| $\omega_{011}$ (cm$^{-1}$) | 200.1 | 194.2 | 191.8 |
| $\gamma_{11}$ (cm$^{-1}$) | 7.1 | 11.7 | 19.0 |
| $\Delta\epsilon_{12}$ | 0.053 | 0.10 | 0.083 |
| $\omega_{012}$ (cm$^{-1}$) | 212.0 | 205.76 | 202.8 |
| $\gamma_{12}$ (cm$^{-1}$) | 5.8 | 11.6 | 14.2 |
| $\Delta\epsilon_{13}$ | 0.032 | 0.028 | 0.026 |
| $\omega_{013}$ (cm$^{-1}$) | 222.1 | 217.0 | 213.8 |
| $\gamma_{13}$ (cm$^{-1}$) | 4.6 | 6.1 | 8.5 |
| $\Delta\epsilon_{14}$ | 0.20 | -- | -- |
| $\omega_{014}$ (cm$^{-1}$) | 238.9 | -- | -- |
| $\gamma_{14}$ (cm$^{-1}$) | 23.5 | -- | -- |
| $\Delta\epsilon_{15}$ | 0.014 | 0.22 | 0.33 |
| $\omega_{015}$ (cm$^{-1}$) | 242.0 | 232.5 | 228.7 |
| $\gamma_{15}$ (cm$^{-1}$) | 3.1 | 18.7 | 27.7 |
| $\Delta\epsilon_{16}$ | 1.03 | 1.43 | 2.31 |
| $\omega_{016}$ (cm$^{-1}$) | 280.7 | 280.7 | 283.3 |
| $\gamma_{16}$ (cm$^{-1}$) | 10.1 | 16.1 | 31.7 |
| $\Delta\epsilon_{17}$ | 0.60 | 0.83 | 0.39 |
| $\omega_{017}$ (cm$^{-1}$) | 294.4 | 291.4 | 299.1 |
| $\gamma_{17}$ (cm$^{-1}$) | 9.7 | 18.9 | 24.5 |
| $\Delta\epsilon_{18}$ | 0.15 | 0.096 | 0.29 |
| $\omega_{018}$ (cm$^{-1}$) | 303.1 | 302.3 | 309.8 |
| $\gamma_{18}$ (cm$^{-1}$) | 9.5 | 11.2 | 22.9 |
| $\Delta\epsilon_{19}$ | 0.13 | -- | -- |
| $\omega_{019}$ (cm$^{-1}$) | 310.4 | -- | -- |
| $\gamma_{19}$ (cm$^{-1}$) | 8.9 | -- | -- |
| $\Delta\epsilon_{20}$ | 0.47 | 0.64 | 0.008 |
| $\omega_{020}$ (cm$^{-1}$) | 316.9 | 312.0 | 322.6 |
| $\gamma_{20}$ (cm$^{-1}$) | 9.5 | 19.1 | 10.2 |
| $\Delta\epsilon_{21}$ | 0.21 | 0.057 | 0.0018 |
| $\omega_{021}$ (cm$^{-1}$) | 343.9 | 340.0 | 337.4 |
| $\gamma_{21}$ (cm$^{-1}$) | 10.2 | 16.2 | 3.4 |
| $\Delta\epsilon_{22}$ | 0.028 | -- | -- |
| $\omega_{022}$ (cm$^{-1}$) | 355.4 | -- | -- |
| $\gamma_{22}$ (cm$^{-1}$) | 4.6 | -- | -- |
| $\Delta\epsilon_{23}$ | 0.028 | 0.41 | 0.30 |
| $\omega_{023}$ (cm$^{-1}$) | 357.8 | 359.6 | 359.7 |
| $\gamma_{23}$ (cm$^{-1}$) | 5.2 | 20.7 | 28.5 |
| $\Delta\epsilon_{24}$ | 0.13 | 0.24 | 0.062 |
| $\omega_{024}$ (cm$^{-1}$) | 362.2 | 377.3 | 367.3 |
| $\gamma_{24}$ (cm$^{-1}$) | 7.6 | 16.1 | 12.5 |
| $\Delta\epsilon_{25}$ | 0.10 | 0.068 | 0.41 |
| $\omega_{025}$ (cm$^{-1}$) | 371.4 | 368.4 | 378.4 |
| $\gamma_{25}$ (cm$^{-1}$) | 7.3 | 16.8 | 23.4 |
| $\Delta\epsilon_{26}$ | 0.33 | 0.20 | 0.019 |
| $\omega_{026}$ (cm$^{-1}$) | 382.5 | 382.3 | 390.6 |
| $\gamma_{26}$ (cm$^{-1}$) | 12.0 | 17.7 | 10.9 |
| $\Delta\epsilon_{27}$ | 0.022 | 0.0042 | 0.018 |
| $\omega_{027}$ (cm$^{-1}$) | 407.3 | 398.2 | 372.9 |
| $\gamma_{27}$ (cm$^{-1}$) | 6.3 | 5.6 | 7.3 |
| $\Delta\epsilon_{28}$ | 0.12 | 0.28 | 0.33 |
| $\omega_{028}$ (cm$^{-1}$) | 418.8 | 414.1 | 399.4 |
| $\gamma_{28}$ (cm$^{-1}$) | 8.4 | 26.7 | 47.5 |
| $\Delta\epsilon_{29}$ | 0.15 | -- | -- |
| $\omega_{029}$ (cm$^{-1}$) | 428.0 | -- | -- |
| $\gamma_{29}$ (cm$^{-1}$) | 11.3 | -- | -- |
| $\Delta\epsilon_{30}$ | 0.039 | -- | -- |
| $\omega_{030}$ (cm$^{-1}$) | 441.4 | -- | -- |
| $\gamma_{30}$ (cm$^{-1}$) | 9.0 | -- | -- |
| $\Delta\epsilon_{31}$ | 0.13 | -- | -- |
| $\omega_{031}$ (cm$^{-1}$) | 462.6 | -- | -- |
| $\gamma_{31}$ (cm$^{-1}$) | 17.4 | -- | -- |
| $\Delta\epsilon_{32}$ | 0.38 | 0.34 | 0.008 |
| $\omega_{032}$ (cm$^{-1}$) | 479.5 | 467.8 | 436.4 |
| $\gamma_{32}$ (cm$^{-1}$) | 14.0 | 50.6 | 10.6 |
| $\Delta\epsilon_{33}$ | 0.70 | 0.52 | 0.21 |
| $\omega_{033}$ (cm$^{-1}$) | 506.3 | 499.5 | 493.2 |
| $\gamma_{33}$ (cm$^{-1}$) | 14.1 | 17.1 | 22.0 |
| $\Delta\epsilon_{34}$ | 0.22 | -- | -- |
| $\omega_{034}$ (cm$^{-1}$) | 516.5 | -- | -- |
| $\gamma_{34}$ (cm$^{-1}$) | 13.0 | -- | -- |
| $\Delta\epsilon_{35}$ | 0.54 | 1.96 | 2.24 |
| $\omega_{035}$ (cm$^{-1}$) | 523.9 | 513.7 | 510.4 |
| $\gamma_{35}$ (cm$^{-1}$) | 12.9 | 32.2 | 43.4 |
| $\Delta\epsilon_{36}$ | 0.30 | 0.37 | 0.56 |
| $\omega_{036}$ (cm$^{-1}$) | 540.0 | 536.5 | 533.4 |
| $\gamma_{36}$ (cm$^{-1}$) | 14.4 | 17.2 | 33.9 |
| $\Delta\epsilon_{37}$ | 0.12 | -- | -- |
| $\omega_{037}$ (cm$^{-1}$) | 549.5 | -- | -- |
| $\gamma_{37}$ (cm$^{-1}$) | 8.9 | -- | -- |
| $\Delta\epsilon_{38}$ | 0.20 | 0.52 | 0.34 |
| $\omega_{038}$ (cm$^{-1}$) | 560.8 | 551.8 | 550.3 |
| $\gamma_{38}$ (cm$^{-1}$) | 12.9 | 31.5 | 45.2 |
| $\Delta\epsilon_{39}$ | 0.10 | -- | -- |
| $\omega_{039}$ (cm$^{-1}$) | 576.0 | -- | -- |
| $\gamma_{39}$ (cm$^{-1}$) | 25.4 | -- | -- |
| $\Delta\epsilon_{40}$ | 0.028 | 0.058 | 0.087 |
| $\omega_{040}$ (cm$^{-1}$) | 600.5 | 592.1 | 588.4 |
| $\gamma_{40}$ (cm$^{-1}$) | 12.6 | 24.2 | 42.1 |
| $\Delta\epsilon_{41}$ | 0.017 | 0.23 | 0.19 |
| $\omega_{041}$ (cm$^{-1}$) | 632.0 | 671.8 | 636.6 |
| $\gamma_{41}$ (cm$^{-1}$) | 14.8 | 86.8 | 89.9 |